\documentclass[10pt,journal,comsoc]{IEEEtran}

\IEEEoverridecommandlockouts
\usepackage{cite}
\usepackage{amsmath,amssymb,amsfonts}
\usepackage{algorithm}  
\usepackage{algpseudocode}
\usepackage{graphicx}

\usepackage{textcomp}
\usepackage{xcolor}
\usepackage{bbm}
\usepackage{bbold}

\DeclareMathOperator*{\argmax}{argmax}

\begin{document}
\title{Deep Learning for Fast and Reliable Initial Access in AI-Driven 6G mmWave Networks 
\thanks{Tarun S. Cousik, Vijay K. Shah, and Jeffrey H. Reed are with Wireless@VT, Bradley Department of Electrical and Computer Engineering, Virginia Tech, Blacksburg, VA 24061, USA. Tugba Erpek and Yalin E. Sagduyu are with Intelligent Automation Inc., Rockville, MD 20878, USA. Emails: \{tarunsc, vijays, reedjh\}@vt.edu, \{terpek, ysagduyu\}@i-a-i.com.}
\thanks{This effort was supported by the U.S. Army Research Office under contract W911NF-20-P0035. The content of the information does not necessarily reflect the position or the policy of the U.S. Government, and no official endorsement should be inferred.}}

\author{\IEEEauthorblockN{Tarun S. Cousik, Vijay K. Shah, Tugba Erpek, Yalin E. Sagduyu, and Jeffrey H. Reed}}

\IEEEtitleabstractindextext{
\begin{abstract}
We present DeepIA, a deep neural network (DNN) framework for enabling fast and reliable initial access for AI-driven beyond 5G and 6G millimeter (mmWave) networks.
DeepIA reduces the beam sweep time compared to a conventional exhaustive search-based IA process by utilizing only a subset of the available beams. DeepIA maps received signal strengths (RSSs) obtained from a subset of beams to the beam that is best oriented to the receiver. In both line  of  sight (LoS)  and  non-line  of sight  (NLoS) conditions, DeepIA reduces the IA time and outperforms the conventional IA's beam prediction accuracy. We show that the beam prediction accuracy of DeepIA saturates with the number of beams used for IA and depends on the particular selection of the beams. In LoS conditions, the selection of the beams is consequential and improves the accuracy by up to $70\%$. In NLoS situations, it improves accuracy by up to $35\%$. 
We find that, averaging multiple RSS snapshots further reduces the number of beams needed and achieves more than $95\%$ accuracy in both LoS and NLoS conditions. Finally, we evaluate the beam prediction time of DeepIA through embedded hardware implementation and show the improvement over the conventional beam sweeping.
\end{abstract}

}

\maketitle
\IEEEdisplaynontitleabstractindextext
\IEEEpeerreviewmaketitle

\vspace{1cm}
\IEEEraisesectionheading{\section{Introduction}\label{sec:introduction}}
A shift to millimeter wave (mmWave) frequencies has been proposed to satisfy the ever increasing bandwidth requirements of wireless communication systems. This shift has forced the adoption of highly directional antennas and/or arrays (HDAs) to combat the path loss associated with these high frequencies. Initial access (IA) in mmWave systems serves the purpose of orienting the HDAs of two or more radio devices (which are unaware of their relative positions) to point at one another to establish the initial connection \cite{giordani2016initial, barati2016initial, alkhateeb2017initial, li2017design}. IA is a key component in 5G and beyond systems to establish the initial connection of a mobile user and the cellular network. 

The adoption of HDAs introduces an alignment window challenge in IA for 5G systems which are currently being deployed and for anticipated 6G systems. This challenge is further exacerbated due to the potential operation at high frequencies such as mmWave bands. Beam alignment becomes more difficult when narrower beams are used at higher frequencies and needs to be repeated more frequently due to potential blockage effects that may otherwise result in loss of beam alignment. IA is an expensive process in terms of delay, transmit power and computational costs involved, and if not properly orchestrated, the communications system may fail to achieve high data rates. 

 5G NR allocates up to a maximum of $64$ synchronization signal blocks (SSBs) for each beam sweep in the mmWave bands. These SSBs contain the relevant information required to establish synchronization between devices. The time allocated to sweep 64 beams is $5$ ms and the beam sweeping process occurs periodically every $5$-$20$ ms \cite{tripathi, waveform}. Each SSB is typically mapped onto a unique beam (sector) in the spatial domain. Consequently, this fixes the number of beams that we need to evaluate before choosing an optimal beam. However, the coherence time (over which the channel remains constant) experienced by mobile platforms is much smaller than $20$ ms. Therefore, novel methods are needed to decrease the time it takes for the IA and improve the connection time to start communications in 5G and beyond. 

The IA time consists of two components, (i) time for beam sweeping where the received signal strengths (RSSs) for different beams are measured, and (ii) time for beam prediction where the beam for a given transmitter-receiver pair to communicate with is identified. The beam sweeping time dominates the overall IA time, and therefore it is essential to improve the IA time by utilizing fewer beams. Transmit power consumption also decreases by utilizing fewer beams in IA. However, conventional beam sweeping (CBS) approaches sweep all beams in general. If they are extended to select the best beam based on RSSs from a reduced subset of beams, this selection may not be necessarily accurate since CBS cannot predict the correct beams for users located in beams that are not present in the subset that is swept. Hence, the CBS approach cannot realize the advantages that arise from using a subset of beams. In this work, exhaustive beam search is referred to as CBS. In general, the mapping from the RSSs measured for a subset of beams to the best beam is a complex process due to various channel, antenna, and network topology effects. This limitation motivates us to understand if and how a data-driven approach could be leveraged to predict best beams that include ones that are not part of the beam sweeping process.          

Machine learning provides automated means to learn from spectrum data and perform complex tasks such as spectrum sensing \cite{davaslioglu2018generative}, signal classification \cite{shi2019deep}, anti-jamming \cite{erpek2018deep}, and waveform design \cite{o2017physical}.  Supported by recent advances in algorithmic techniques and computational resources, deep learning has emerged as a viable solution to capture high-dimensional representations of spectrum data \cite{erpek2020deep}. The `mappable' nature of input RSS to beams is particularly well suited for a deep neural network (DNN)-based approach. In this paper, we present DeepIA, a deep learning-based solution for IA that reduces the beam sweep time by utilizing measured RSSs from a subset of all available beams and maps them to the best oriented beam in the entire set of beams.

The DNN trained in DeepIA learns to associate the correct/best oriented beam  between the transmitter and receiver with the RSS measured by the receiver. Hyperparameters of the DNN are carefully selected to avoid both underfitting and overfitting. For both line-of-sight (LoS) and non-line-of-sight (NLoS) mmWave channel conditions, we compare CBS and DeepIA using various number of beams and show that, DeepIA not only provides a faster IA scheme but also a more reliable one in terms of prediction accuracy. For example, DeepIA predicts the optimal beam with close to $95\%$ accuracy in LoS conditions by sweeping only $7$ out of $24$ beams. The accuracy of the CBS for the same setting is limited to $24\%$. While a CBS approach would traditionally not use a subset of beams, nevertheless, we implement it in this fashion to achieve a fair comparison between the accuracy and input size of the two approaches. 

We identify that the DNN's accuracy is not just dependent on the number of beams used in its input, but is also dependent on the specific combination of beams that are used. In other words, if we fix the number of inputs to the DNN's, certain combinations of beams are more effective in accurately predicting the best oriented beams. We propose using a sequential feature selection (SFS) algorithm to help identify better performing combinations of beams for DeepIA's input. We find that using beams predicted by the SFS algorithm further appreciably improves the accuracy of DeepIA in the LoS case. In LOS conditions, it offers enhanced and robust improvement in accuracy for situations where $8$ or fewer beams are utilized. Not much improvement is observed by SFS in NLoS cases. Motivated by the performance gain obtained by feature selection in LoS situation, we investigate alternate solutions to improve the NLoS performance and find that using the averaged RSSs as input further improves the performance in both LoS and NLoS conditions. For instance, a DeepIA network that uses the averaged inputs from $25$ measurements is able to achieve a $95\%$ accuracy with just $7$ beams in NLoS conditions. 

We validate the practicality of our approach by  computing the time needed to run DeepIA's DNN through embedded implementation with the Field Programmable Gate Array (FPGA) and comparing it against the computation time for the CBS. We find that both computations are on the order of a few microseconds. The beam sweep time is on the order of milliseconds for CBS and a few hundred microseconds for DeepIA depending on the channel conditions. This analysis supports the claim that DeepIA decreases the overall time required for initial access.

DeepIA is poised to reduce the overall operational costs as it reduces the number of transmissions/beams each transmitter has to sweep through during IA. Consequently, it also decreases the temporal and spatial footprint of the signal over the air which reduces the interference from/to other ongoing transmissions (such as in spectrum sharing scenarios) as well as the probability of detection/intercept. Thus, DeepIA could potentially improve the resiliency against out-of-network interference and jamming.

The rest of the paper is organized as follows. 
Section \ref{sec:relatedwork} discusses related work. Section \ref{sec:systemmodel} describes the system model and the assumptions. Section \ref{sec:proposed_solution} describes in detail the DeepIA framework. Section \ref{sec:perfanalysis} provides an analysis on prediction accuracy and computation time taken by CBS and DeepIA in LoS and NLoS mmWave channel models. Section \ref{sec:conclusion} concludes the paper. A preliminary version of the material in this paper is available in \cite{cousik2020fast}.   

\section{Related Work} \label{sec:relatedwork}
Conventional IA involves an exhaustive beam search. A predefined number of beams divide the azimuth plane into sectors, with each beam covering a unique sector\cite{wei2018initial, wei2017exhaustive, jasim2017simultaneous}. A common implementation of the exhaustive search involves the transmitter cycling through every sector while the receiver runs in a quasi-omnidirectional mode and listens for the transmitted signal. The receiver (e.g., a user equipment (UE) in 5G) records the RSS  for each transmit beam and finally sends back the beam with the highest RSS to the transmitter (e.g., gNodeB in 5G). The transmitter then turns on this particular beam, and the receiver performs the same beam sweeping process from its side to determine its best sector. This exhaustive search is computationally inefficient. Unlike omnidirectional sub-6 GHz systems, this IA process has to be periodically repeated in mmWave systems in order to ensure that there is no misalignment between transmitter-receiver pairs over time. 

Another variant of the exhaustive search runs narrow transmitter beams against narrow receiver beams\cite{wei2018initial}. Iterative search uses a combination of wide and narrow beams on the transmitter and receiver to minimize the computational cost of the exhaustive search at the expense of decreasing the detection accuracy  \cite{wei2017exhaustive}. In addition to this, since only a partial set of the antenna elements are used to create the beams, it can decrease the range over which these wide beams can be used. A hybrid IA process was introduced in \cite{wei2017exhaustive}, where first the iterative process is performed and then the receiver sends the uplink signals in the best beam. The transmitter finds its best narrow beam after cycling through all its narrow beams against the best receiver beam. This hybrid IA process performs equally with the iterative search in terms of accuracy but utilizes approximately $86$-$90\%$ of the time resources. Both the iterative and hybrid approaches are handicapped by the dependency to use two sets of beams. As the number of beams in a system increases, the cost of the memory (typically SRAM based solution) needed to store corresponding antenna element weights increases rapidly, thus discouraging commercial applicability of these types of solutions. Additionally, while they are faster than the exhaustive beam sweep, at their very best they tend to perform as well as exhaustive beam search \cite{wei2017exhaustive}.

Using multi-beam analog beamformers which simultaneously overlay several narrow beams in the spatial domain for IA was proposed in \cite{jasim2017simultaneous}. While the transmit time is reduced, computational and temporal requirements for the back-end processing needed to distinguish the beams and detection accuracy remain unclear. \cite{Fastlink} proposed the use of compressed sensing to leverage the change gradient of RSS/SNR to detect the best beam from a subset of beams. They found that their approach scaled appreciably with an increase in the sparsity of a channel. \cite{RapidIA} leveraged the DeepIA algorithm as a means of determining the action-policy in a reinforcement learning framework called DeepIA-DRL. Additionally, they also propose a new algorithm called RapidIA, that attains beam orientation with a single transmission. This is achieved by leveraging transmissions from/to several coordinated base stations/APs to transmit to the user which roughly translates to transmitting multiple beams but from different spatial regions. It shows good promise in scenarios whenever multiple BS can exist within each other's broadcasting range. The subsequent related work that are discussed leverage various forms of context-based information to enhance and accelerate the IA process. The average discovery time by leveraging knowledge obtained from real time arrival statistics of incoming users was reduced in  \cite{soleimani2019fast}.     
Gated recurrent neural networks (RNNs) were used in \cite{mazin2018accelerating} to predict the sequence of the beams swept during IA. Using call records to find the user's location, \cite{mazin2018accelerating} used this information to determine the number of users in each sector such that the beams are swept in a descending order of users in each sector, i.e., sector with the highest number of users goes first and so on. Pseudo-directional patterns were considered in \cite{mazin2018accelerating} since the operations are in the sub-6 GHz band.  
Leveraging channel state information (CSI) as the input variable for a DNN was investigated in \cite{sim2020deep} to predict the $k$ best beams for initial access.  
A context-aware IA approach was proposed in \cite{zia2019machine}, where global positioning system (GPS) data is fed to the UE and BS to help predict the optimal beam choice for IA. The main challenge for this approach is to generate the reliable GPS data for various scenarios including indoors. 

In our work, we compare DeepIA results with the CBS since it is the de-facto technique 5G NR prescribes and show the performance improvement in terms of time and accuracy. The simple nature of collecting and processing RSS makes it feasible for DeepIA to be implemented on hardware platforms.  
Iterative  and hybrid techniques do not necessarily perform better than CBS; as a result, an increase in DeepIA's accuracy over CBS also means DeepIA's accuracy is higher than both hybrid and iterative techniques.

\section{System Model} \label{sec:systemmodel}
\begin{figure}[h!]
	\centerline{\includegraphics[width=0.7\linewidth]{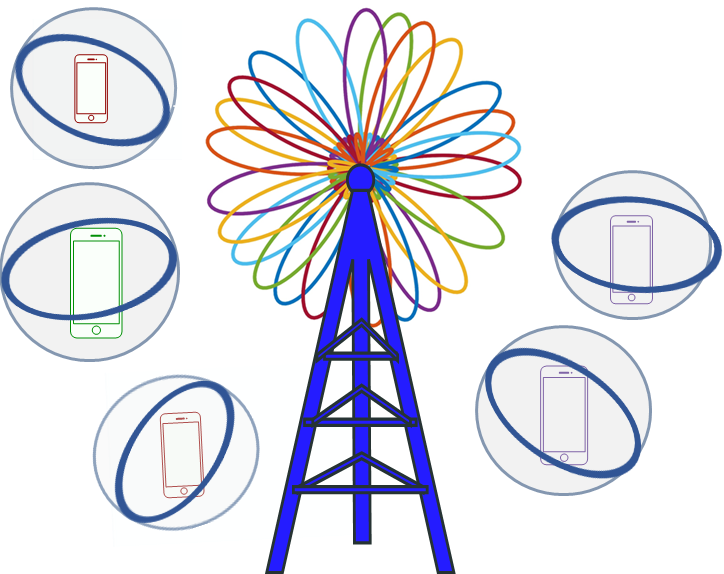}}
	\caption{System model.}
	\label{fig:sys}
\end{figure}

We consider a 6G mmWave network that consists of a directional transmitter and omnidirectional receivers as shown in Fig. \ref{fig:sys}. Without loss of generality, we consider a 2D plane where $R$ receivers are uniformly distributed in a square area and the transmitter is located at the center. 
No receiver is placed within a $1$ meter radius of the transmitter to support the close intercept (CI) model explained in Section \ref{subsec:mmWaveCha}. First, the transmitter performs the IA process. While it is not explicitly shown, it is assumed that the subsequent processes in this paper are repeated again from the receiver's side as is the case with any IA strategy outlined in the related work. 

\subsection{Transmitter and Antenna Array Characteristics}
\begin{figure}[h!]
\begin{center}
\includegraphics[scale=0.55]{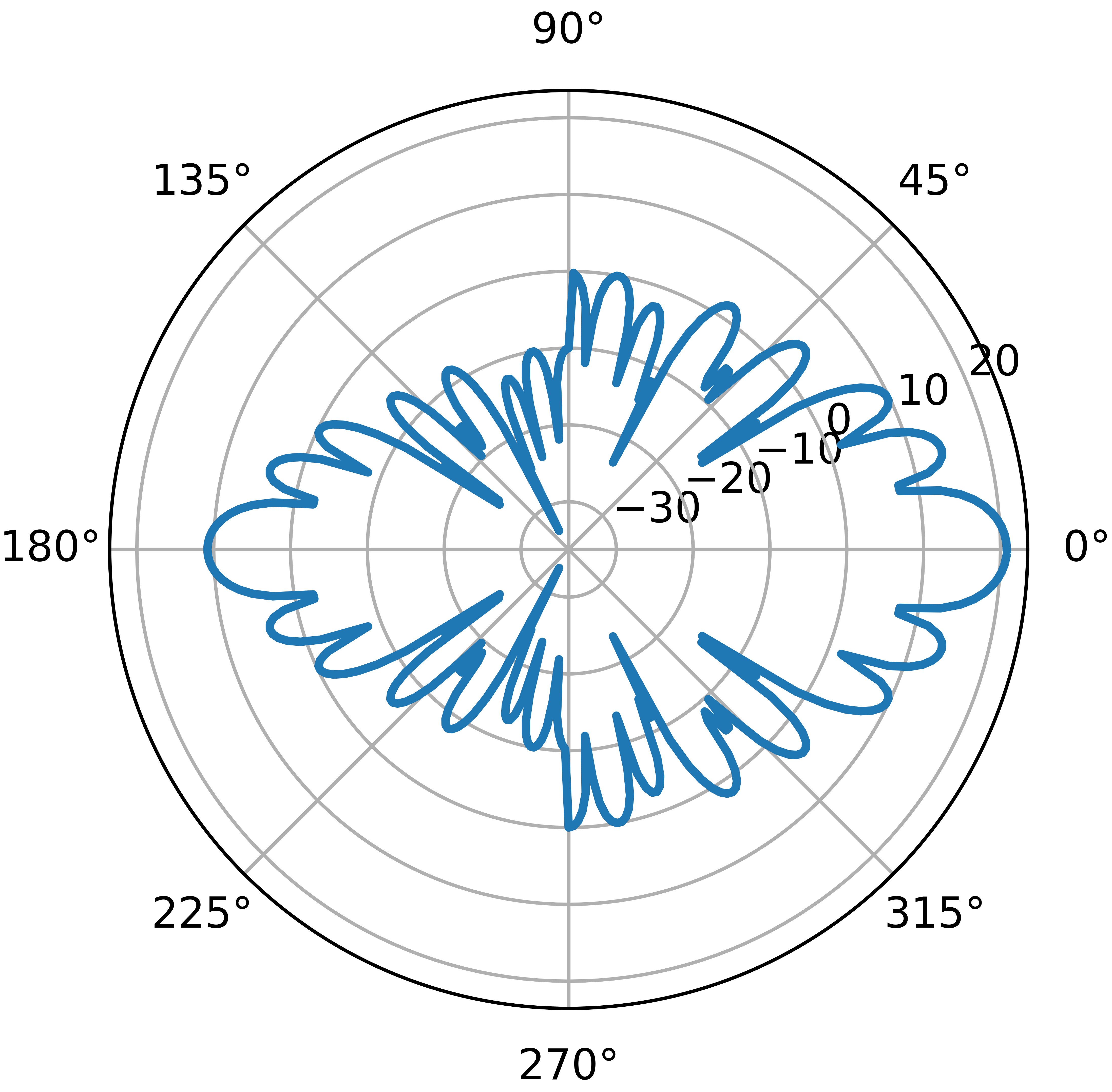}
\caption{Antenna pattern (gain values in dB).}
\label{fig:AP}
\end{center}
\end{figure}

We design a $10 \times 10$ antenna array at the transmitter which generates a beam width of about $15 ^{\circ}$ using the standard planar array formulation \cite{balanis2016antenna}. This array points at an azimuth angle $\phi = 0^{\circ}$ and an elevation angle $\theta = 90^{\circ}$. Since we consider a 2-D scenario, we take the azimuthal slice of the array factor at $\theta = 90^{\circ}$. We then subtract $10$ dB from this standard array slice for azimuth angles between $180^{\circ}$ and $360^{\circ}$ as shown in Fig. \ref{fig:AP}. This modification serves two purposes. First, antenna designers routinely modify the array/antenna patterns to ensure minimal backlobe radiation. The goal is to avoid transmitting in unnecessary directions, reduce electromagnetic interference to back-end electronics and avoid interfering with users in the backlobe's direction.  
Second, using a standard $10 \times 10$ planar array pattern severely degrades the performance of CBS because of pattern symmetricity. When the array factor of the backlobe has the same magnitude as that of the front lobe, the CBS process gets confused on which beam to pick.  
This  modified array pattern is consistently used for all of the sectors.

\subsection{Channel Models} \label{subsec:mmWaveCha} 
In this work, we consider both LoS and NLoS mmWave channels. Specifically, we use the close-in (CI) path loss model, where at distance $d$, the path loss in dB is given by 
\begin{equation}
 PL(d) = PL(d_{0}) + 10 n \log_{10}\frac{d}{d_{0}} + X_{\sigma}, 
\end{equation}
where $PL(d_{0}) = 10\log_{10}\left(\frac{4 \pi d_{0}}{\lambda}\right)^2$, $\lambda$ is the signal wavelength, $d_0 = 1$ m, $n$ is the best fit minimum mean squared error path loss exponent (PLE) (note $n = 4.5$ for NLoS conditions and $n = 1.9$ for LoS conditions) and $X_{\sigma}$ is the shadow factor representing large scale signal fluctuations (note $X_{\sigma}$ is a zero-mean Gaussian random variable with a standard deviation of $10$ for NLoS conditions and $1.1$ for LoS conditions \cite{rappaport2015wideband}). 

The LoS and NLoS parameters in \cite{rappaport2015wideband} were obtained from empirical results, thus meriting applicability in practical situations.  
Other alternative path loss models derived from ray tracing simulations have also been presented in the literature. While these ray tracing models offer a complex environmental model, the models they deliver are specific to the locations and situations they are simulated for. Additionally, these models are better suited for experiments that require additional parameters such as number of reflections and delay characteristics.  

\subsection{Overview of the Baseline IA} \label{subsec:baseline}

The baseline IA that we refer to is Conventional Beam Sweeping (CBS) approach that uses exhaustive search (i.e., sweep all beams \cite{wei2018initial}).  
We define $\mathcal{N} = \{1,...,N\}$ as the set of all $N$ beams that the transmitter can sweep. 
Let $r_k$ denote the $k$th receiver, where $k\in\{1,2,3,...,R\}$, and $RSS_{ik}$ denote the RSS corresponding to the $i$th beam between the transmitter and the receiver $r_k$. In CBS, all $N$ beams are swept sequentially and $\hat{i}_k$th beam which provides the highest RSS is selected, i.e., for receiver $r_k$, the selected beam is $\hat{i}_k = \argmax\limits_{i \in \mathcal{N}} RSS_{ik}$. Algorithm \ref{P0_algo} shows the baseline CBS algorithm.

\begin{algorithm} [h!]
\label{alg_1}
	\textbf{Input:} Receiver $r_k$\\
	\textbf{Output:} Selected beam $\hat{i}_k$
	\begin{algorithmic}[1]
	\For{$i \in \mathcal{N}$} 
	\State Measure $RSS_{ik}$
	\EndFor
	\State $\hat{i}_k = \argmax\limits_{i \in \mathcal{N}} RSS_{ik}$
	\end{algorithmic}  
	\caption{Baseline CBS Algorithm}
	\label{P0_algo}  
\end{algorithm}	

\subsection{Overview of DeepIA}\label{subsec:overview_DeepIA}

While a detailed description of DeepIA's inner workings is provided in Section \ref{sec:proposed_solution}, it is beneficial to know that DeepIA uses the global information of receiver positions during the training phase. Given that the receiver's position is known to the transmitter during the training stage, the best beam is calculated based on the angle between them; this process is described in Section \ref{subsec: beam_mapping}. The DNN is trained by feeding the RSS values from a subset of beams, $\mathcal{M}$ (where $M$ is the cardinality, i.e., $M = |\mathcal{M}|$), as the input, where $\mathcal{M} \subseteq \mathcal{N}$.
 
To support fast implementation on an embedded platform (see Section~\ref{subsec:timeAnalysis}), the DNN uses a simple feedforward neural network (FNN) architecture. The input layer consists of $M$ neurons, each receiving input from one beam from $\mathcal{M}$.  
At the output layer, there are $N$ output neurons corresponding to the $N$ beams and their magnitudes denote the respective probabilities of being the optimal beam. During training, the label, i.e., the best beam for a given transmitter and receiver position, is determined based on the angle between the transmitter and receiver instead of the measured RSS values. The subsets of $M$ beams are uniformly sampled from $N$ and the same subset is used during both the training and test times in our simulations.

\section{Deep Learning Framework of DeepIA} \label{sec:proposed_solution}
This section describes the details of DeepIA's deep learning framework; namely, DNN architecture, beam mapping and label generation, and training and testing phases. 
\subsection{Deep Neural Network Architecture} \label{subsec:NNarchitecture}

The DNN architecture of DeepIA consists of seven layers including the input and output layers. The input layer has $M$ neurons, and the output layer has $N$ neurons with each neuron providing a likelihood score. The hidden layers have $32$, $64$, $128$, $64$ and $32$ neurons, respectively. These hidden layer structures are determined after tuning the hyperparameters based on the training and validation performance. In simulations, we set $N =24$ and select $M$ between $1$ and $24$. 
All the hidden layers are activated by a ReLu activation function and Softmax activation function is used at the output layer. The number of neurons in each layer and the activation functions used are shown in Table \ref{table:DeepIANNArch}.
Each layer's output except that of the output layer is batch normalized before passing it on to the next layer. This architecture was selected for after comprehensive experimentation with various other architectures, after the validation accuracy showed that the other models either over-fit or under-trained the training data.

\begin{table}[h!]
\caption{Deep Neural Network Architecture.}
\centering
{\small
\begin{tabular}{l|l|l}
\textbf{Layers} & \textbf{Number of neurons} & \textbf{Activation function}\\ \hline  \hline
Input & $M = 1,2,3, \dots,$ or $24$ &  $-$  \\ \hline 
Dense 1 & 32 & ReLu \\ \hline
Dense 2 & 64 & ReLu \\ \hline
Dense 3 & 128 & ReLu \\ \hline
Dense 4 & 64 & ReLu \\ \hline
Dense 5 & 32 & ReLu \\ \hline
Output  & $N = 24$ & Softmax \\
\end{tabular}
}
\label{table:DeepIANNArch}
\end{table}

\subsection{Beam Mapping and Label Generation} \label{subsec: beam_mapping}

\begin{figure}[h!]
\begin{center}
\includegraphics[scale=0.45]{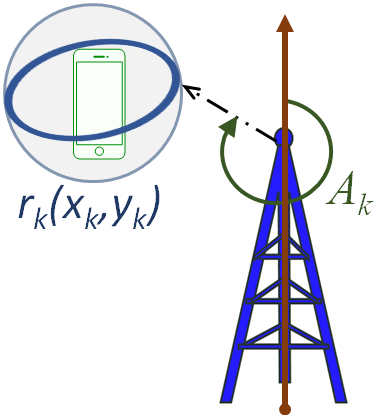}
\caption{Angle between the transmitter and the receiver.}
\label{fig:angle}
\end{center}
\end{figure}

As shown in Fig. \ref{fig:angle}, the angle between the $k$th receiver ${r_k}$ and the transmitter is denoted as $A_k$. This angle represents the relative orientation between any receiver position and transmitter. $A_k$ can take values between $0^{\circ}$ and $359.99 ^{\circ}$, and is rounded to the lowest integer. Transmitter $A_k$ for every receiver $r_k$ is then mapped to its corresponding beam as follows. When $N= 24$, angles $((-8^{\circ}, 7^{\circ}],(7^{\circ},22^{\circ}],(22^{\circ},37^{\circ}],\dots,(337^{\circ},352^{\circ}])$ are mapped to Beams $1, 2, ..., 24$ respectively. Note that $-a^{\circ}$ is the same as $360^{\circ}-a^{\circ}$. This is performed by setting $i^*_k$ =  $\left\lceil\frac{A_k +8}{15}\right\rceil$, where $i^*_k$ represents the true beam sector that receiver $r_k$ should correspond to. $i^*_k$ also corresponds to the labels that we use against our training, test and validation features. In addition, we use $i^*_k$ in evaluating the accuracy of both the CBS and DeepIA. 
In practical situations, wireless network providers can obtain these labels  from propagation modelling tools that model the localized environment and/or through measurement campaigns in the local area. 

\subsection{Deep Neural Network Training} \label{subsec:training} 

DeepIA uses RSS values from $M$ beams as inputs. The typical training time varies depending on the number of beams, number of epochs as well as the hardware used in training. In simulations, we use GoogleColab Pro to train the DNNs and for any given input of $M$ beams, it takes roughly 2 minutes to train a DNN. Since training will be performed offline, the training time and computational resources can be ignored in subsequent performance comparisons.  
The input variables are transformed to linear units and normalized using the maximum of the absolute value. Note that the values in Tables \ref{tab:trainingdata} and \ref{tab:testingdata} are only written in dB scale, to aid in visualizing the procedure. In reality, the DNN has input features that range from $0$ to $1$.
The DNN is trained to learn to map $\left\{RSS_{ik}\right\}_{i \in \mathcal{M}}$ values to the corresponding $i^*_k$ values. 
We show an example of how DNN training set would look like in Table \ref{tab:trainingdata}. The receiver column corresponds to the index of different receivers. The true beam column corresponds to the best beam that is determined based  on  the  angle  between  the  transmitter  and  receiver instead of the measured RSS values. 

We separate $R$ receivers and their corresponding data into $3$ sets for training, validation and testing. The number of receivers used for training, validation and testing are given by $R_{tr}$, $R_{val}$ and $R_{te}$, respectively. Therefore,  $R = R_{tr} + R_{val} + R_{te}$. The output of training for beam set $\mathcal{M}$ is the trained DNN model $\texttt{DeepIA}_{\mathcal{M}}: \left\{ RSS_{ik} \right\}_{i \in \mathcal{M}} \rightarrow \hat{i}_k \in \mathcal{N}$. 
\begin{table}[h!]
    \centering
    \caption{An example of training data entries.}
    \begin{tabular}{c|c|c|c|c}
         Receiver & Beam $1$ & \dots & Beam $M$ & True Beam \\ \hline \hline
         1 & -30 & \dots & -46 & 1 \\ \hline
         2 & -51 & \dots & -24 & 5 \\ \hline
         - & - & \dots & - & - \\ \hline
         $R_{tr}$ & -65 & \dots & -68 & 2 \\ 
    \end{tabular}
    \label{tab:trainingdata}
\end{table}

DNN models are trained for $M = 1,2\dots24$ and $10$ epochs are used to train each model. The backpropagation algorithm is used to train the DNN by minimizing the cross entropy loss function. We experimented with various learning rates and finally chose an initial learning rate of $10^{-2}$  and a final learning rate is $0.2$. A batch size of $1024$ is consistently maintained for all the models. We observed that while the training time for smaller batch sizes  (i.e., $512,128,64$) took longer they offered no benefit in terms of validation accuracy. Larger batch sizes (i.e., $5000,10000$) showed slight deterioration in performances and we settled for $1024$. The hyperparameters of the DNN are determined after evaluating the validation accuracy and epochs where the training loss saturated. After checking a variety of optimizers including ADAM, Adamax, Stochastic Gradient Descent, and Adabound, our results indicate that Adabound converges to the minimum of the loss function the quickest, while the accuracy is practically identical for Adabound and Adam. Therefore, the Adabound optimizer \cite{luo2019adaptive} is used during training.
 
\subsection{Deep Neural Network Testing} \label{subsec:testing}
The RSS values from the same $M$ beams used during training are measured and fed to the DNN model for receiver indices that have not been encountered in the training process. The index of the output with the highest probability is selected as the best spatially oriented beam. This process is shown in Algorithm \ref{P0_algo2}. Table~\ref{tab:testingdata} shows an example test input and output with $\mathcal{M}$ inputs. The predicted beam should match exactly with the true beam for a correct prediction. In cases where they do not match up such as that of receiver $r_{k}$, DeepIA makes an error in its prediction.

\begin{algorithm} [h!]
\label{alg_2}
	\textbf{Input:} Receiver $r_k$, $\texttt{DeepIA}_{\mathcal{M}}$\\
	\textbf{Output:} Predicted beam $\hat{i}_k$
	\begin{algorithmic}[1]
	\For{$i \in \mathcal{M}$}
	\State Measure $RSS_{ik}$
	\EndFor
	\State $\hat{i}_k = \texttt{DeepIA}_{\mathcal{M}}\left(\{RSS_{ik}\}_{i \in \mathcal{M} }\right)$
	\end{algorithmic}  
	\caption{DeepIA Algorithm}
	\label{P0_algo2}  
\end{algorithm}	

\begin{table}[h!]
    \centering
    \caption{An example of test data entries.}
    \begin{tabular}{c|c|c|c|c|c}
         Receiver & Beam $1$ & \dots & Beam $M$ & Predicted & True \\ 
         & & & & Beam & Beam \\\hline \hline
         1 & -35 & \dots & -46 & 1 & 1 \\ \hline
         2 & -51 & \dots & -64 & 5 & 5 \\ \hline
         - & - & \dots & - & - \\ \hline
         $r_{k}$& -27 & \dots & -71 & 4 & 5\\ \hline
         - & - & \dots & - & - \\ \hline
         $R_{te}$ & -25 & \dots & -28 & 2 & 2 \\ 
    \end{tabular}
    \label{tab:testingdata}
\end{table}

\subsection{Selection of Features} \label{sec:featuresel}
As discussed above, the input to the DeepIA framework is the subset of beams, $\mathcal{M}$. Thus, an important question arises at this point on the selection of the beams in the subset $\mathcal{M}$. We investigate whether certain beams matter more than others when it comes to the actual selection of the beams. 

In order to investigate this question, for any subset $\mathcal{M}$ of beams, we utilize a baseline set that consists of $M$ beams that are spatially equally separated. We call this set as Manually Selected Beams (MSB) denoted by $\mathcal{M}_{MSB}$.
The MSBs for a given number of beams $M$, can be calculated as
\begin{equation} \label{eq:MSB}
\mathcal{M}_{MSB} (j) =\left\{ \left \lfloor j\frac{24}{M} \right \rfloor \bigg| \:j\in \left\{0,1,2,\dots,M-1 \right\} \right\},
\end{equation}
where $\lfloor \cdot \rfloor$ is the floor function and  $\mathcal{M}_{MSB} (j)$ denotes the $j$th beam in the set $\mathcal{M}_{MSB}$.

It is important to identify the best combination of input beams to increase prediction accuracy while minimizing the total number of beams used. In fact, the whole domain of feature selection is developed with the objective of identifying a minimal set of features for a given dataset, that when used, typically minimizes the objective function. Feature selection can be categorized into three basic approaches:

\textit{Filter Methods}: These methods rely on processing input variables with respect to some criterion (e.g., mutual information, Pearson's correlation, Fischer score, etc.), ranking the processed data and then selecting the $M$ desired number of input variables that obtain the highest $M$ ranks \cite{FS_survey}.

\textit{Embedded Methods}: These methods incorporate the feature selection process as part of the  training process; regularization can in essence be thought of as feature selection. Typically algorithms like Lasso/L1 norm based regularization introduce feature sparsity while approaches like Ridge regression help regulate performance in the presence of large outliers\cite{L1L2}.

\textit{Wrapper Methods}: Wrapper methods typically involve cycling through the various combination of input variables in the learning algorithm in question, measuring a predetermined score and then selecting the feature set with the best score. This technique is favorable because the features selected are specific to both the input data and learning algorithm.

 Note that, since the total combinatorial search space for $24$ input variables is over $16\times10^{6}$, exhaustively cycling through all of them is not practical. To overcome this limitation, in this work, we utilized a variant of the wrapper method termed the naive sequential feature selection (SFS) \cite{FS_survey}. DeepIA served as the learning algorithm for the SFS, the prediction accuracy obtained from the validation data served as the score. SFS reduces the feature search space to $299$ combinations. On average, it takes about $2$ minutes to train DeepIA for a given number of beams and this wrapper-based feature selection algorithm takes on average $30$ hours in total for any given channel condition. 
 
We show the implementation of the SFS approach for selecting optimal $\mathcal{M}$ subset of beams (out of $\mathcal{N}$ beams) in Algorithm \ref{algo3}.
 As shown in line 1, an empty beam subset $\mathcal{M}$ is initialized. Then, beams (or features) are iteratively added to the subset $\mathcal{M}$, on the condition that the beam that is included obtains the highest prediction accuracy amongst all the other available beams (See lines 4 - 14). Beams that are selected and added based on these criteria  are permanently retained in the $\mathcal{M}$ and this process is repeated until the required  number of features $(M)$ are added (See line 2).
 
\begin{algorithm} 
	\textbf{Input:} Number of beams to be used, $M$  \\
	\textbf{Output:} Selected subset of beams, $\mathcal{M}$
	\begin{algorithmic}[1]
	\State Chosen beam subset, $\mathcal{M} = \phi$
	\For{$i = 1, \dots, M$}
	    \State $Acc_{max} = -1$
	    \For{beam, $b \in N \setminus \mathcal{M}$}
	        \State $\mathcal{M}_{temp} = \mathcal{M} \cup b$
	        \State $\hat{i}_k$ =  DeepIA$_{\mathcal{M}_{temp}}$ ($\{RSS_{ik}\}_{i \in \mathcal{M}_{temp}}$)
	        \State $Acc_{temp} = \frac{\sum_{k=1}^{R_{val}} \mathbbm{1}( i^*_k = \hat{i}_{k})}{R_{val}}$ 
	        \If{$Acc_{temp} > Acc_{max}$}
	            \State $Acc_{max} = Acc_{temp}$
	            \State $\mathcal{M} = \mathcal{M}_{temp}$
	        \EndIf
	        \State $\mathcal{M}_{temp} = \mathcal{M}_{temp} \setminus b$
	    \EndFor
	\EndFor
	\State return $\mathcal{M}$
	\end{algorithmic}
	\caption{{Sequential Feature Selection (SFS) Approach for selecting $\mathcal{M}$ subset of beams}}
	\label{algo3}

\end{algorithm}

\section{Performance Analysis} \label{sec:perfanalysis}

\subsection{Simulation Setting}

 We consider a two-dimensional 6G mmWave network scenario where transmitters and receivers all lie on the same plane. The location of the transmitter is fixed at $(0,0)$ and the receiver's X and Y positions each uniformly randomly distributed between $-25$ m and $25$ m. This bounds the simulation cell to an area of $50 \times 50$ m. The transmit power is set to $20$ dBm.  
 
 A total of $10^6$ receiver positions are generated, which act as data samples. Out of total $10^6$ data samples, $65\%$ are used to train DeepIA, $15\%$ of data samples are used for validation, and $20\%$ of data samples are used for testing.

\subsection{Comparison Approaches}
We evaluate the performance of DeepIA and compare it to that of CBS in terms of both accuracy and beam prediction time, when varying the subset of beams out of total beams.

\noindent \textbf{Prediction Accuracy:}   
    The prediction accuracy is computed as the ratio of receivers for which the predicted beam matches the true beam, namely
    \begin{equation}
        \text{Prediction Accuracy} [\%] = \frac{\sum_{k=1}^{R_{te}} \mathbbm{1}( i^*_k = \hat{i}_{k})}{R_{te}} \times 100,
    \end{equation}
    where $\mathbbm{1}$ is the indicator function ($\mathbbm{1}(E) =1$ if $E$ holds and $0$, otherwise).

\noindent \textbf{IA Time:}  The IA time is the total time incurred in predicting the correct beam for a certain transmitter-receiver position. It constitutes two parts described below. 
    \begin{enumerate}
        \item \textit{Beam Sweep Time --}  the time taken to sweep a desired number of beams. The beam sweep time is directly proportional to the number of beams swept. For instance, 5G NR \cite{tripathi} allows a duration of one half frame or $5$ ms for sweeping across a total of $64$ beams. Using these numbers, it should take $1.875$ ms to sweep $24$ beams, $0.9375$ for $12$ beams,  $0.675$ ms for $8$ beams, and so on. In simulations, we consider the same subset of beams to sweep for both DeepIA and CBS. Therefore, we compare the accuracy performance under the same beam sweep time.   
        
        \item \textit{Beam Prediction Time --} the time it takes to process the collected RSS data and predict the beam it belongs to. In CBS, this roughly involves calculating a maximum operation on the RSS collected from $N$ beams. In DeepIA, this involves calculating the time it takes to run the DNN model using $M$ inputs in order to predict the beam a given receiver belongs to. 
    \end{enumerate}

In this section, we compare the performance of DeepIA against CBS, in terms of prediction accuracy and computational time. The prediction accuracy portion of the work, also examines the improvement in accuracy obtained when the SFS beams are used in both LoS and NLoS channel conditions.  
We then analyze performance improvement observed when the average RSS of multiple measurements are fed to the DeepIA.

In the time analysis section, we discuss the benefits of utilizing DeepIA in terms of time taken for computation.

\subsection{Prediction Accuracy Analysis}
\begin{figure}[ht]
    \centering
\includegraphics[scale = 0.55]{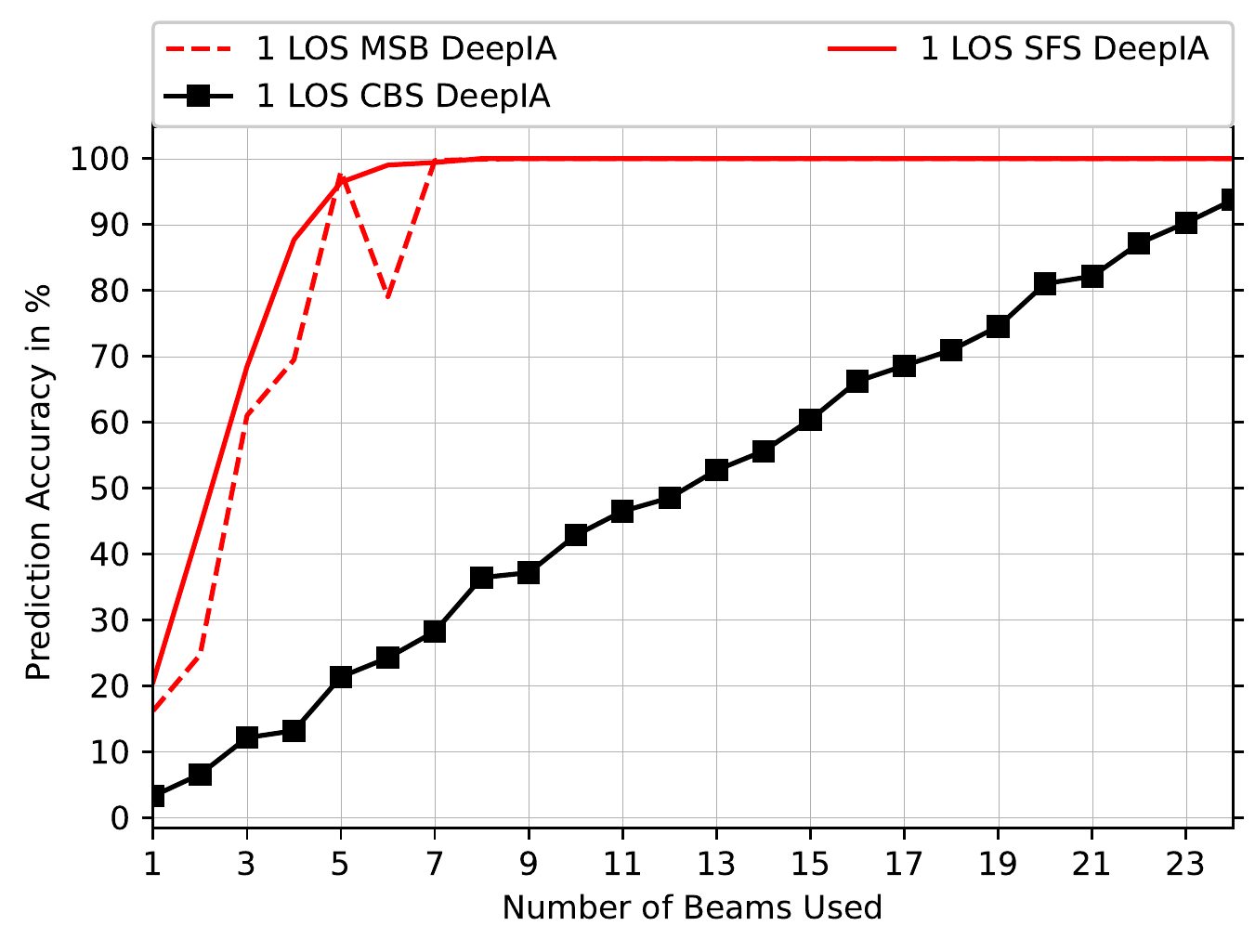}
\caption{Improvement from using Feature Selection in LoS.}
\label{fig:LOS_results}
\end{figure}

Since the exhaustive beam search in CBS relies on scanning the entire spatial plane, its accuracy continuously deteriorates as fewer beams are utilized. The user distribution contributes to the slightly non-linear prediction accuracy curve seen in Fig. \ref{fig:LOS_results}. In addition, the shadow fading factor in LoS condition also causes incorrect prediction in a few users (explained in later part of this section), thus accuracy is $95\%$ with 24 beams as shown in Fig. \ref{fig:LOS_results}. On the other hand, in LoS conditions DeepIA demonstrates a more robust and reliable prediction accuracy over the range of beams that are swept as shown in Fig. \ref{fig:LOS_results}. For example, in the LoS scenario, setting $M=7$ achieves close to $100\%$ prediction accuracy. The performance of DeepIA deteriorates when less than $7$ beams are utilized. 

\begin{figure}[ht]
    \centering
\includegraphics[scale = 0.55]{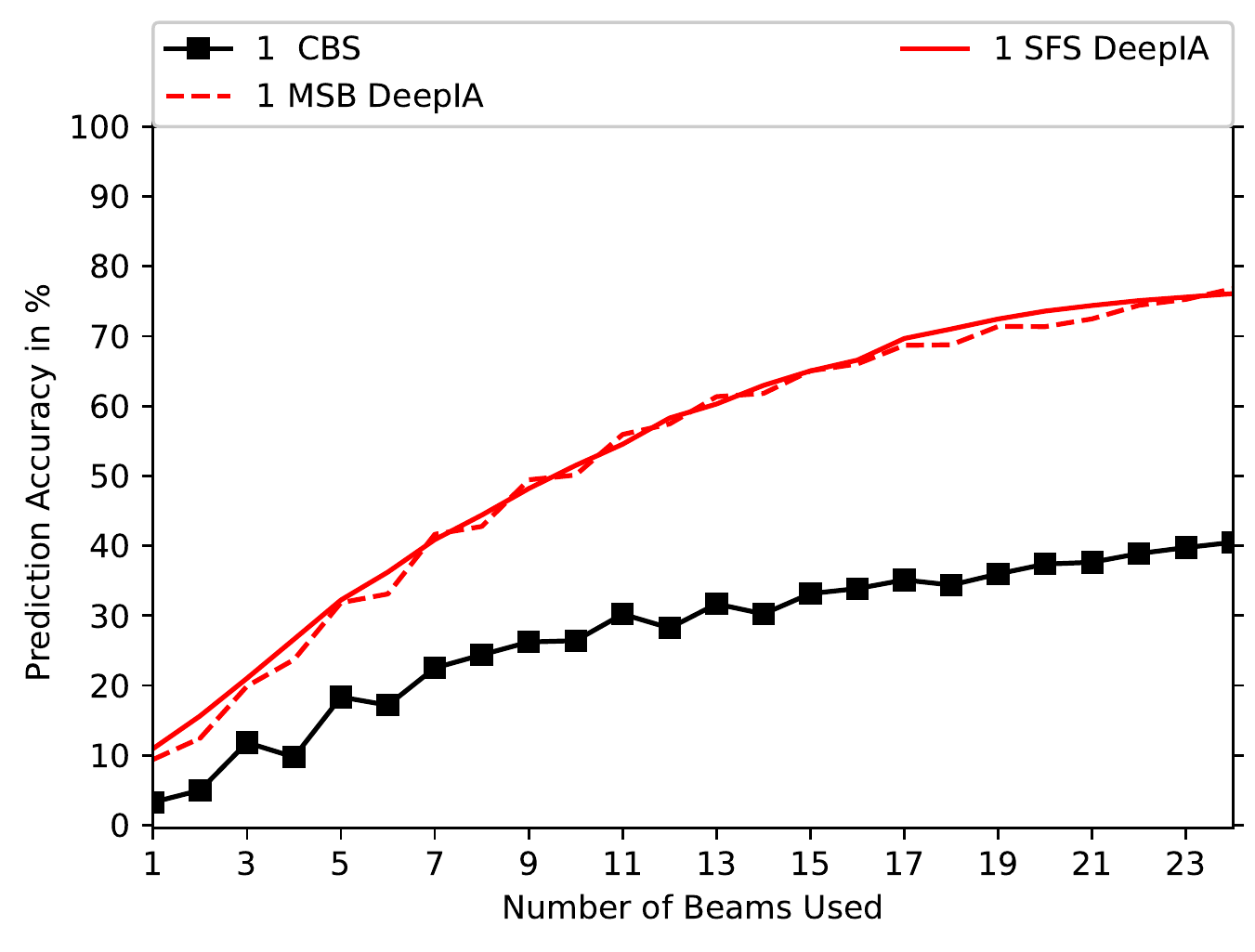}
\caption{Improvement from using Feature Selection in NLoS.}
\label{fig:NLOS_results}
\end{figure}

The relevance of feature selection becomes evident as we observe that the prediction accuracy by DeepIA is better with a $5$ beam input than a $6$ beam input, when these input beams are selected based on the MSB criterion. Recall, MSB refers to the Manually Selected Beams which are equally separated from one another as described in  Eq. \ref{eq:MSB}. If each feature/beam is considered equally important, this finding is obviously counter-intuitive. In other words, DeepIA is able to perform better with $5$ beams selected using the MSB criterion, because the `quality of the information' in those beams is more valuable to DeepIA  in determining the prediction accuracy. When beams are selected based on the SFS algorithm, we find that the prediction accuracy for DeepIA models that take inputs of $1$ to $6$ beams are all further improved relative to the MSB case. Furthermore, we note that the beams selected based on the SFS criterion correspond to a rough upperbound to what the MSB can achieve. The maximum accuracy achieved with a $24$ beam CBS in LoS conditions is achieved using a DeepIA model trained on just $5$ SFS selected beams. In LoS scenario, SFS is able to reasonably discern `quality' beams for DeepIA. Therefore, using more than $8$ adds very little benefit in terms of accuracy in the DNN approach under LoS conditions.    

With the introduction of the severe shadow fading in the NLoS mmWave channels, DeepIA's performance deteriorates compared to LoS case as shown in Figure \ref{fig:NLOS_results}. Nonetheless, DeepIA still outperforms CBS. In fact, the gap in performance increases as the number of input beams increases from $1$ to $24$. The  NLoS fading factor adds 
considerable amount of fluctuations in  the  collected RSS values and this in effect leads to the following two phenomena which contribute to the rapid drop in CBS accuracy under NLoS conditions. 

Since the fluctuations are random, it can increase or decrease the value of the `clean' RSS  measurements where `clean' refers to the calculated RSS values based on the path loss for a given distance excluding the shadow fading factor. The clean RSS values from users positioned in between the main lobes of two adjacent beams decreases rapidly as the user position changes closer to the boundary/intersection of the two beams. Fluctuations due to shadowing 
can cause the RSS of the incorrect adjacent beam to exceed the RSS of the correct beam,  
thus resulting in inaccurate predictions for the CBS. These fluctuations directly impact the performance of CBS for both LoS and NLoS conditions by boosting the RSS of an incorrect beam.  
Moreover, in NLoS conditions, the relatively large shadow fading factor is able to influence the back lobe and force it to overshoot the main lobe's power, thus adding another possible culprit that reduces prediction accuracy in CBS.  

Fig. \ref{fig:NLOS_results} also shows that the introduction of sequential feature selection (SFS) does not improve the overall performance by reasonable measure, but the earlier observation we made on the SFS-based DeepIA being the upper limit of the MSB-based DeepIA still holds. 
Previous work \cite{reunanen2003overfitting} which implemented  sequential feature on various other data sets reported that SFS could potentially have higher prediction accuracies for a model trained with lower number of input features and attributed this effect to the forced retention of features in the SFS approach. In this paper, we did not observe that effect and conclude that this effect may be data dependent. 

In certain time-constraint cases, it would be practical for LoS scenarios to use a feature selected DeepIA with $7$ beams, since it gives close to $100\%$ and performs better than an exhaustive search with $24$ beams in CBS while utilizing approximately $29\%$ of the total computation time. If the performance is critical and the channel is predominantly NLoS, it makes sense to use DeepIA with $24$ beams.
Thus, DeepIA ordinarily provides the network operator with a prediction accuracy vs. beam alignment time trade off to work with. However, in next subsection, we discuss an additional ``averaging'' enhancement which allows performing IA with just a subset of beams even in NLoS conditions; thus, realizing faster and reliable IA in both LoS and NLoS conditions.

\subsection{Averaging of Input Beams and its Effects on Accuracy} \label{subsec: averaging}
It is clear that the larger NLoS  shadow fading factor has a significant impact on the performance regardless of the number of beams used. Motivated by the fact that the LoS performance far exceeds the NLoS performance, we consider alternative ways to boost the NLoS accuracy. It is expected that DeepIA in NLoS may benefit from input data fed with less variance,  
similar to what it got in LoS conditions. 
Since averaging a signal decreases the variance  
we build a new dataset that is of size $[R, M, s]$ where $s$ is the number of measurements/snapshots collected for averaging. Recall, $R$ and $M$ are the number of users and the number of beams in the subset, respectively. We select the values of $s$ as $10$ and $25$. Averaging provides a significant increase in the performance of DeepIA for NLoS conditions. We observe that in NLoS conditions, DeepIA converges to a more than $95\%$ prediction accuracy with just $6$ SFS beams if $25$ averages are used (and with $11$ SFS beams if $10$ averages are used). This leads to an interesting observation for our dataset -  when fed with data corrupted by AWGN noise, even if the DeepIA network leverages the higher quality features it tends to perform at the level of the MSB. In other words, while noise brings down the performance of a network fed with MSBs, it tends to affect a DeepIA network fed with SFS features more severely. The lower limit seems to plateau towards the MSB performance. 

\begin{figure}[ht]
    \centering
\includegraphics[scale = 0.55]{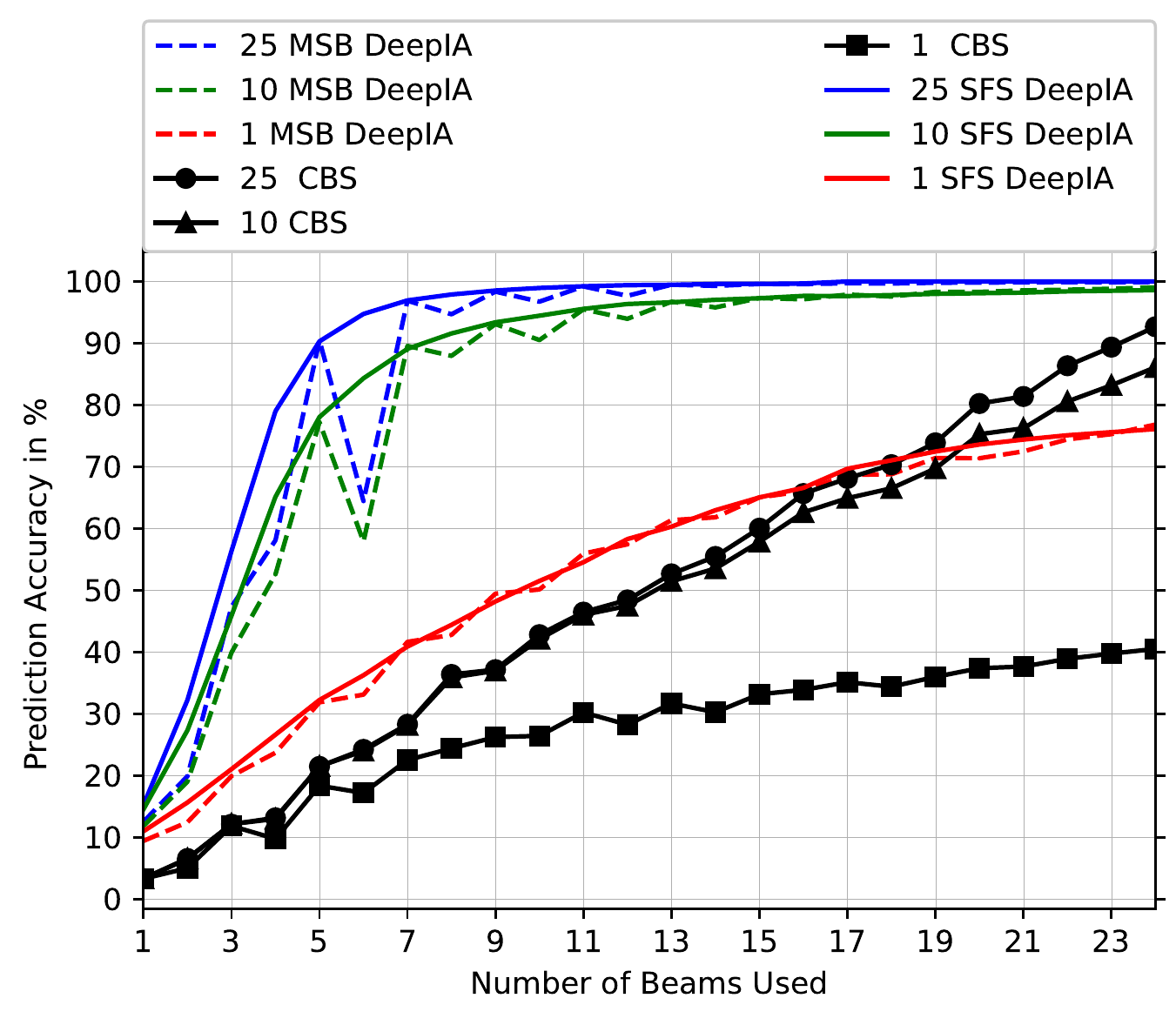}
\caption{Improvement from using averaging data in NLoS.}
\label{fig:avg_NLOS_results}
\end{figure}

\begin{figure}[ht]
    \centering
\includegraphics[scale = 0.55]{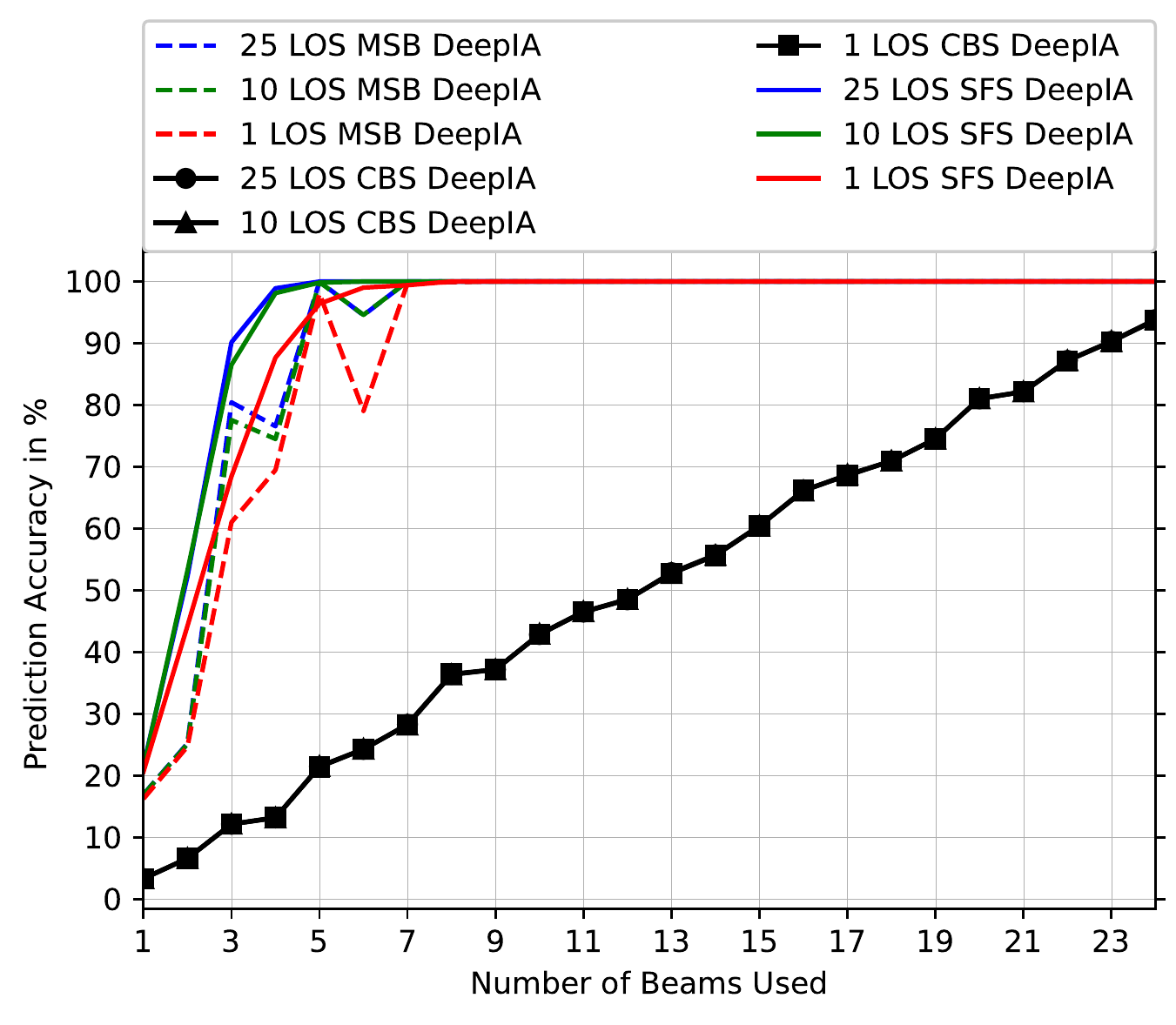}
\caption{Improvement from using averaging data in LoS.}
\label{fig:avg_LOS_results}
\end{figure}

\begin{figure}[h!]
\begin{center}
\includegraphics[scale=0.55]{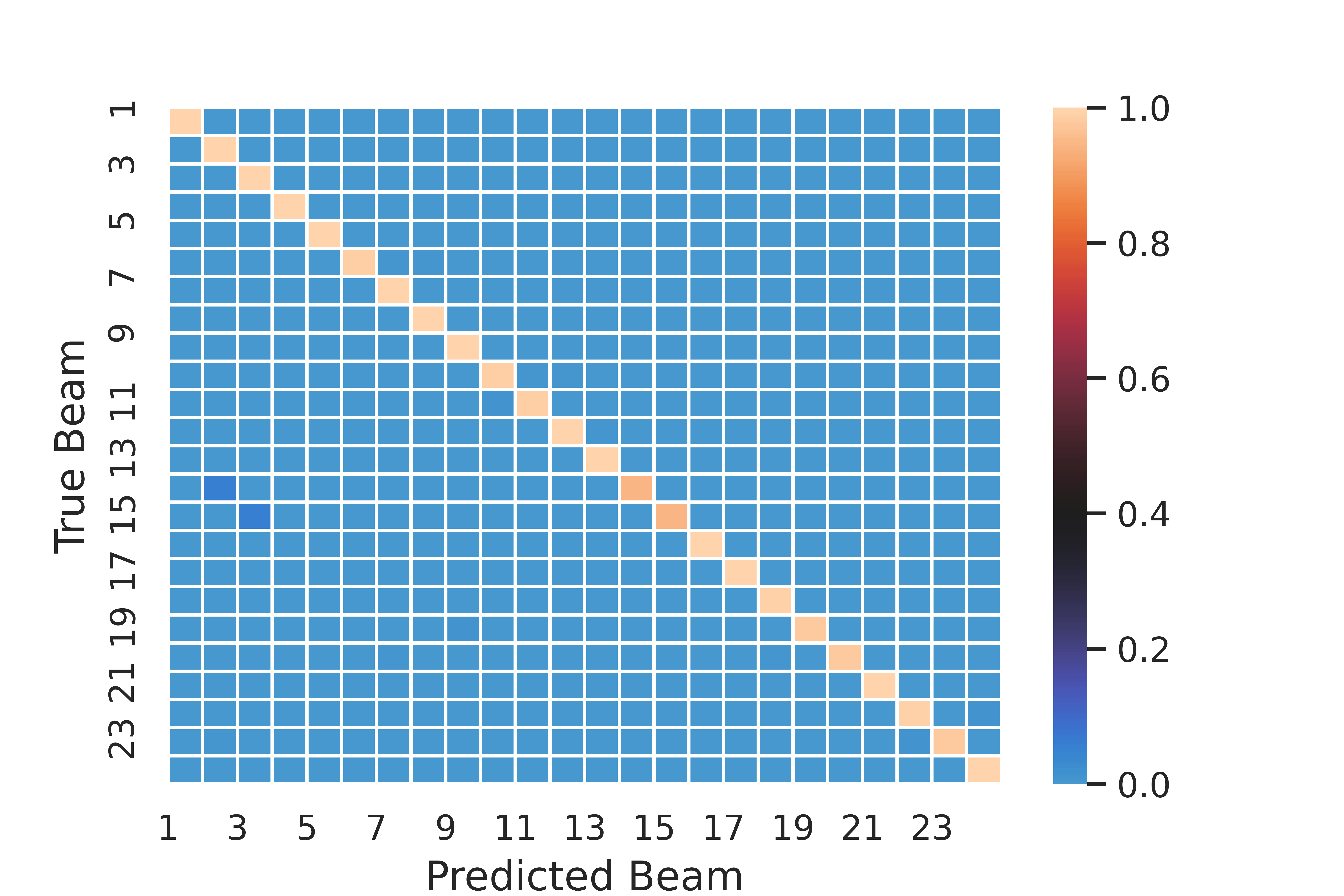}
\caption{Confusion matrix of Beam Predictions from  a $4$ beam DeepIA framework in LoS conditions.} \label{fig:conf_matrix_LoS_4}
\end{center}
\end{figure}

\begin{figure}[h!]
 \centering
\includegraphics[scale=0.55]{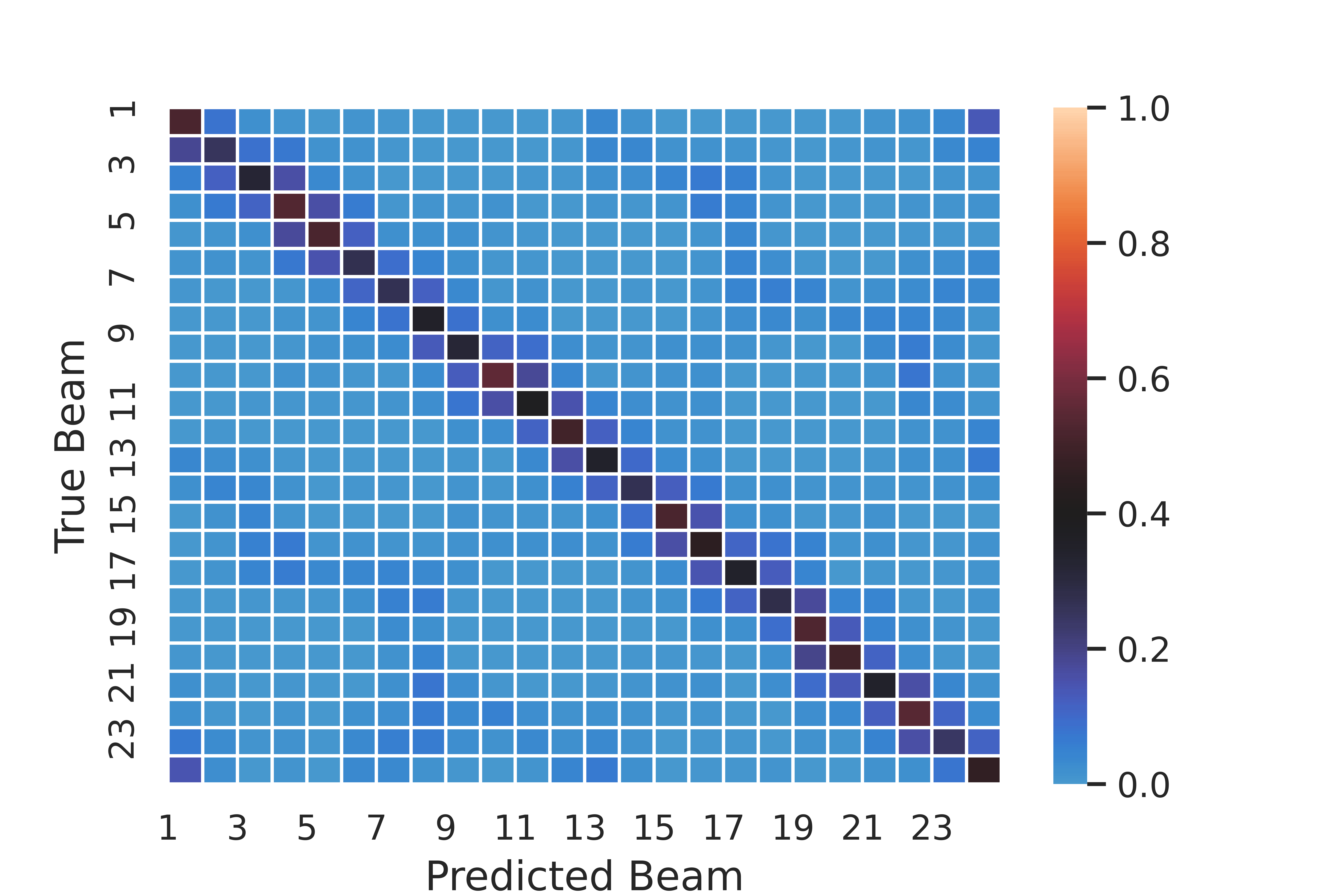}
\caption{Confusion matrix of Beam Predictions from  a $7$ beam DeepIA framework in NLoS conditions.} \label{fig:conf_mat_NLoS_7}
\end{figure}

\begin{figure}[h!]
 \centering
\includegraphics[scale=0.55]{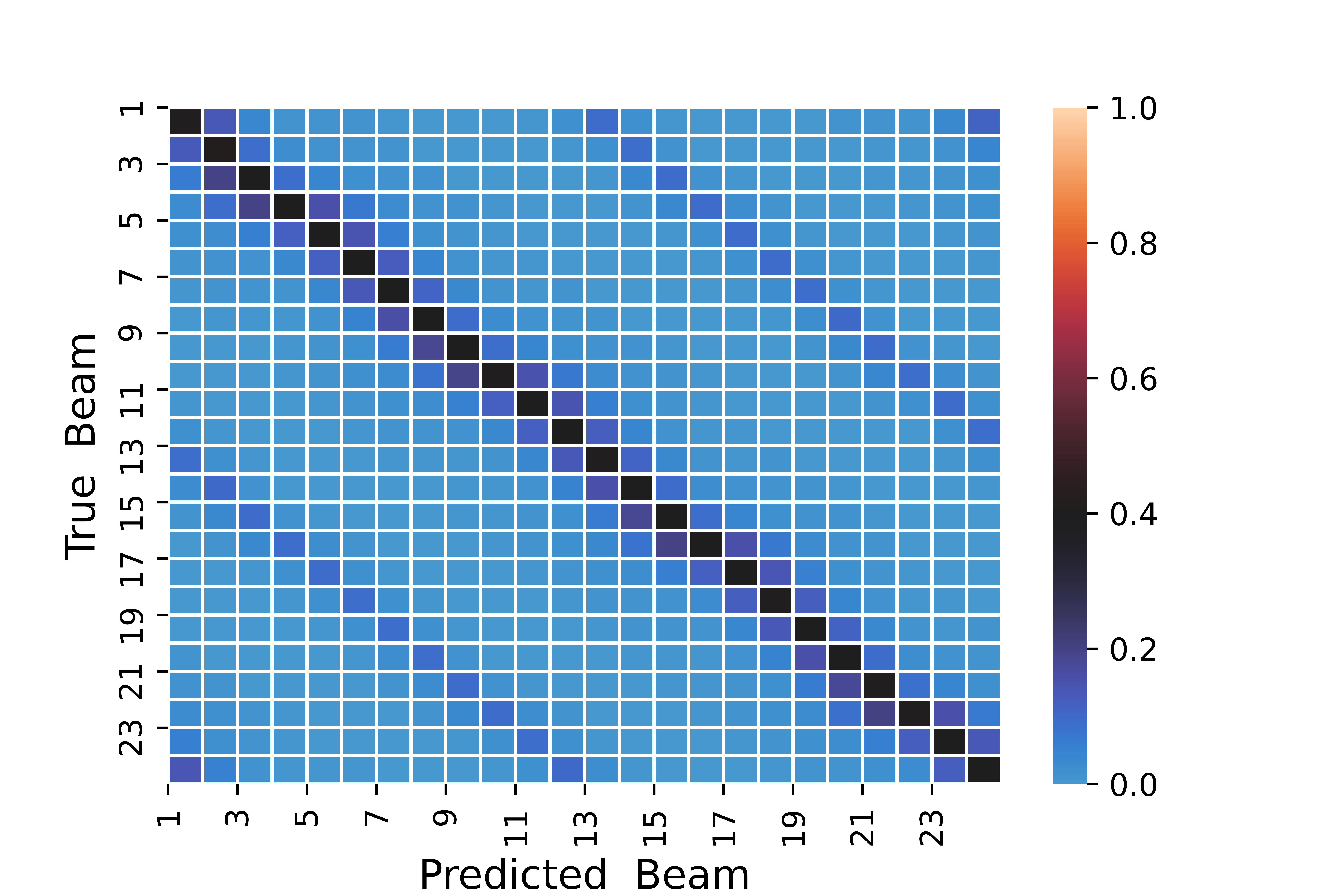}
\caption{Confusion Matrix for CBS with 24 beams in NLoS.} \label{fig:conf_mat_NLos_24}
\end{figure}

\begin{figure}[h!]
 \centering
\includegraphics[scale=0.55]{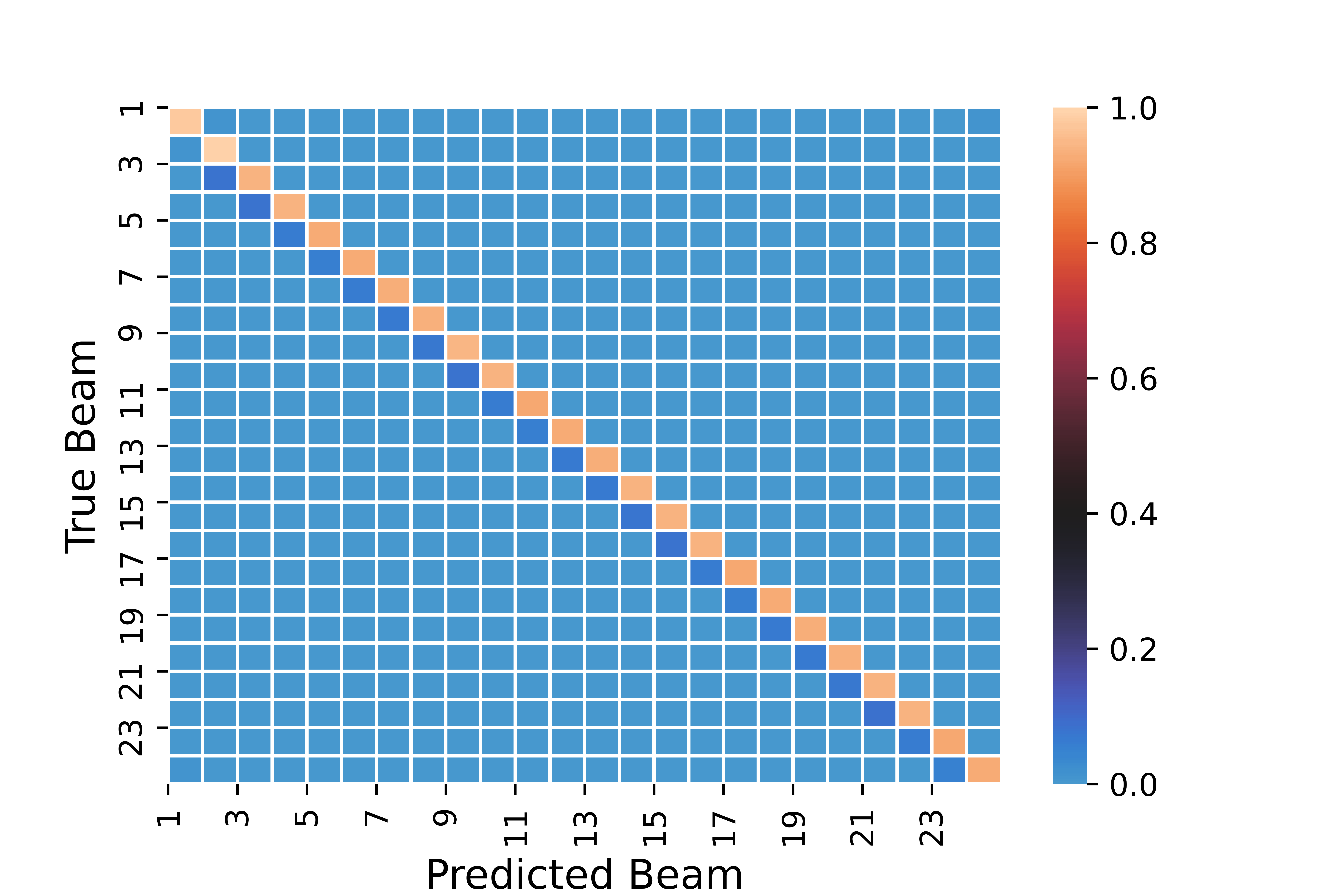}
\caption{Confusion Matrix for CBS with 24 beams in LoS.} \label{fig:conf_mat_LoS_24}
\end{figure}

The confusion matrices in Figures \ref{fig:conf_matrix_LoS_4}, \ref{fig:conf_mat_NLoS_7}, \ref{fig:conf_mat_NLos_24}, and \ref{fig:conf_mat_LoS_24} show how prediction accuracy is affected at the beam level. The previous prediction degradation due to fluctuations distorting adjacent and main beams can be seen in all the CBS confusion matrices. We can also observe that, in the NLoS CBS case, the back lobe begins to degrade prediction accuracy.  
In DeepIA, the deep learning network has the ability to recognize this back lobe related performance degradation and circumvents it by relying on the general trend of RSS values for each beam it observes. Nonetheless, it is still susceptible to the effects of fluctuations and  sometimes incorrectly selects adjacent beams. However, prediction accuracies for the users in individual beams are not uniform and do not seem to be related to the input beams fed to the network.

\subsection{Beam Prediction Time Analysis} \label{subsec:timeAnalysis}
For practical implementation aspects, 5G allows $5$ ms periodicity by default to complete the beam sweeping procedure in the IA. However, coherence time over which the channel remains unchanged is much smaller for highly mobile users communicating at high frequencies. Coherence time $T_C$ is inversely proportional to the maximum Doppler spread $D_S = \frac{f_c v}{c}$, where $f_c$ is the center frequency, $v$ is the speed of mobile user, and $c$ is the speed of light \cite{Tse}. As an example, consider $f_c = 28$ GHz. If the user moves with walking speed (e.g., $v = 1.4$ m/s), then $T_C$ is less than $7.7$ ms. On the other hand, if the user moves with a car speed of $v = 25$ m/s, then $T_C$ is less than $0.43$ ms. Therefore, it is important to make the beam prediction decision within this time frame.

To better understand the computation time in practice, we consider running the DNN configuration of DeepIA on an embedded platform. For that purpose, the procedure of \cite{soltani2019real} to convert the trained software model for a deep neural network to the FPGA code is followed and computational aspects of running DeepIA's DNN are evaluated for the FPGA implementation. Vivado Design Suite \cite{Vivado} is used to simulate and then synthesize the FPGA code with the 16-bit implementation for Xilinx UltraScale FPGA.

We evaluate the processing time associated with beam sweeping and beam prediction as follows. Beam sweeping involves a pilot tone transmitted from each beam at the transmitter one at a time to the receiver. The receiver measures the RSS value for each beam. This process is repeated $N$ times for CBS algorithm and $M$ times for DeepIA. Note $N$ is the total number of beams in consideration, and $M \subseteq N$ is the number of subset of beams using in DeepIA framework. The selection of the best beam from RSS values can either be performed at the transmitter or receiver. In the former case, the receiver transmits the measured RSS value back to the transmitter after every beam transmission as a feedback. In the latter case, the receiver predicts the best beam itself and transmits a single feedback. Assuming it takes $t_s$ to receive a tone signal at the receiver, the beam sweeping time is reduced from $Nt_s$. to $Mt_s$ by switching to DeepIA. This is because the duration of beam sweeping is linearly proportional to the number of beams swept, and we find that it is much larger than the beam prediction time that we discuss next.

The duration of beam prediction, however, depends on whether CBS or DeepIA is used. Suppose $M = 7$ beams are swept to fix the beam sweeping time for both CBS and DeepIA. The duration to take the maximum operation in CBS is $MT$ with one comparator, $ \left(\left\lceil \frac{M}{2}\right\rceil +1\right)T$ with two comparators, and so on, where $T$ is the time for one FPGA cycle at $100$ MHz clock. Hence, CBS selects its beam in at most $0.06$ $\mu$s. On the other hand, DeepIA runs the RSSs from $6$ beams through its  DNN. The latency to process one input sample (namely, $6$ RSSs) through the DNN architecture given in Table \ref{table:DeepIANNArch} is $3.85$ $\mu$s. Since the beam prediction time in both CBS and DeepIA is at most at microsecond level, the time for beam sweeping dominates the total time spent for IA. Therefore, DeepIA reduces the time for IA significantly and leaves more time for data communications to carry ever-growing traffic demands of 5G and beyond.

When considering averaging the RSSI data, it is reasonable to question if there are any additional time costs associated with that. The 5G NR standard specifies a subcarrier spacing of $240$ MHz or a symbol duration of 4.16 microseconds for the mmWave frequency sync signals\cite{mobility}. The bandwidth required for the synch transmission is roughly 60 MHz or a time period of 16 ns. Each synchronization signal block (SSB) is 4 symbols long and the length of 1 symbol is $4.16$ $\mu$s . The SSB comprises of the Primary Broadcast Channel (PBCH), Secondary Synchronization Signal (SSS) and Primary Synchronization Signal (PSS). The SSS is responsible for the RSSI estimation and is one symbol long. Assuming a sampling speed of $5$ GSPS (available from dual channel mmWave ADCs) \cite{TI_DS}, we note that we can obtain $21840$ samples in $1$ symbol/SSS duration or about 870 samples for each RSSI value, assuming $25$ averages are performed within the 1 symbol duration. Note that the number of I/Q samples used for capturing the RSSI data is specific to the chip/radio being implemented. We limit the scope of this work to investigating how the prediction accuracy scales with the number of samples used in averaging. There are other radio design constraints that may dictate the number of  samples that are needed per average, so the basis of this work should not be used to determine that. If the number of averages we need is not served in 1 symbol duration, the averaging would be done over multiple SSS, however this impacts both CBS and DeepIA.   

\section{Conclusion} \label{sec:conclusion}
In this paper, we developed the deep learning-based  IA solution, DeepIA, for 5G and 6G mmWave networks. DeepIA collects measurements only from a small number of beams to predict the best beam for data communications. It is suggested as an alternative to  CBS that iteratively searches for the best beam from among all possible beams. DeepIA can successfully capture the complex patterns of transmitter-receiver locations and beam patterns, and can predict the best beam with very high accuracy even when a small number of beams are swept in the IA. Compared to CBS, DeepIA significantly reduces the time for IA and leaves more time for communications. Our simulation results show that DeepIA is able to predict the optimal beam with nearly $100\%$ accuracy even when $7$ out of $24$ beams are used in LoS conditions, while the beam prediction accuracy of CBS drops to around $24\%$ with the same number of beams utilized. We identified that certain beams are more valuable in determining accuracy and leveraged this to increase prediction accuracy to close to $100\%$ using just 5 beams in LoS. In NLoS we found that the significant amount of fluctuations present prevented significant features from contributing to further increase in accuracy. We also found that by using the averaged RSSI we could further boost the performance of both LoS and NLoS conditions and that NLoS can exceed $95\%$  prediction accuracy. Finally, we assessed the beam prediction time of DeepIA through embedded implementation on FPGA and showed that DeepIA significantly reduces the time needed for IA.  

\bibliography{references.bib}
\bibliographystyle{unsrt}

\end{document}